\documentclass[sn-mathphys,Numbered]{sn-jnl}

\usepackage{graphicx}
\usepackage{multirow}
\usepackage{amsmath,amssymb,amsfonts}
\usepackage{amsthm}
\usepackage{mathrsfs}
\usepackage[title]{appendix}
\usepackage{xcolor}
\usepackage{textcomp}
\usepackage{manyfoot}
\usepackage{booktabs}
\usepackage{algorithm}
\usepackage{algorithmicx}
\usepackage{algpseudocode}
\usepackage{listings}
\begin{document}

\title[Article Title]{Temporal-Spatial Manipulation of Bi-Focal Bi-Chromatic Fields for Terahertz Radiations}

\author[1,4]{\fnm{Jingjing} \sur{Zhao}}
\author*[3]{\fnm{Yizhu} \sur{Zhang}}\email{zhangyizhu@tju.edu.cn}
\author[1]{\fnm{Yanjun} \sur{Gao}}
\author[2]{\fnm{Meng} \sur{Li}}
\author[2]{\fnm{Xiaokun} \sur{Liu}}
\author[2]{\fnm{Weimin} \sur{Liu}}
\author[1]{\fnm{Tian-Min} \sur{Yan}}
\author*[2,1]{\fnm{Yuhai} \sur{Jiang}}\email{jiangyh3@shanghaitech.edu.cn}

\affil[1]{\orgdiv{Shanghai Advanced Research Institute}, \orgname{Chinese Academy of Sciences}, \orgaddress{ \city{Shanghai}, \postcode{201210}, \country{China}}}

\affil[2]{\orgdiv{Center for Transformative Science and School of Physical Science and Technology}, \orgname{ShanghaiTech University}, \orgaddress{\city{Shanghai}, \postcode{201210}, \country{China}}}

\affil[3]{\orgdiv{School of Precision Instrument and Optoelectronics Engineering}, \orgname{Tianjin University}, \orgaddress{\city{Tianjin}, \postcode{300072}, \country{China}}}

\affil[4]{\orgdiv{}\orgname{University of Chinese Academy of Sciences}, \orgaddress{\city{Beijing}, \postcode{100049}, \country{China}}}

\abstract{

Mixing the fundamental ($\omega$) and the second harmonic (2$\omega$) waves in gas phase is a widely employed technique for emitting terahertz (THz) pulses. The THz generation driven by bi-chromatic fields can be described by the photocurrent model, where the THz generation is attributed to free electrons ionized by the $\omega$ field, and the 2$\omega$ field provides a perturbation to break the symmetry of the asymptotic momentum of free electrons. However, we find that the THz radiation is amplified by one order of magnitude when driven by bi-focal bi-chromatic fields, contradicting the common understanding of the photocurrent model. Meanwhile, present measurements demonstrate that the THz radiation mainly originates from the plasma created by the 2$\omega$ pulses instead of the $\omega$ pulses. Energy transfer from the 2$\omega$ beam to the THz beam during the THz generation has been observed, validating the major contribution of the 2$\omega$ beam. Furthermore, the THz bandwidth has been observed to extensively exceed the bandwidth of the pump pulse, not be explained by the photocurrent model as well. These counterintuitive results indicate that undiscovered physical mechanisms are involved in bi-chromatic THz generation in plasma, presenting a significant challenge for understanding strong-field nonlinear optics and simultaneously expanding various applications.}


\maketitle

\section{Introduction}\label{sec1}

The demand for super-continuum terahertz (THz) pulses, capable of encompassing a broader spectrum of fingerprint spectral lines associated with rotational and vibrational resonance transitions, is significant in both commercial and scientific applications of spectroscopic techniques. In comparison to nonlinear optical crystals emitting narrow-band THz pulses \cite{zhang_terahertz_1992,han_use_2000,vicario_generation_2014}, mixing bi-chromatic strong laser fields in a gas-phase medium has emerged as a popular alternative for generating high-intensity super-continuum THz pulses \cite{cook_intense_2000,fulop_laserdriven_2020,liao_perspectives_2023,simpson_spatiotemporal_2024}. The efficiency of optics-to-THz conversion generated from ambient air plasma induced by bi-chromatic fields of 800 nm and 400 nm with a 35 fs pulse duration is approximately 0.01$\%$ \cite{oh_intense_2013}, and the THz bandwidth can be extended up to 40 THz \cite{hah_enhancement_2017}. To enhance the THz conversion efficiency, a longer pump wavelength can be employed \cite{wang_efficient_2011,jang_efficient_2019,koulouklidis_observation_2020,mitrofanov_ultraviolet--millimeter-band_2020,nikolaeva_scaling_2022}, while a broader THz bandwidth can be generated by a few-cycle pump pulse with broader bandwidth \cite{thomson_terahertz_2010,matsubara_ultrabroadband_2012}.

The THz generation from bi-chromatic fields can be explained by the photocurrent model. According to this model, the THz pulse generation inside a plasma primarily results from the transient photocurrent of free electron induced by the femtosecond field \cite{zhang_experimental_2020,zhang_continuum_2020,gao_coulomb_2023,fan_trajectory_2023}. Several commonly accepted perspectives can be derived from the photocurrent model: (1) Free electrons are predominantly generated through ionization by the high-strength $\omega$ pulse. The presence of the 2$\omega$ pulse as a perturbative field creates a symmetry-broken laser field, thereby introducing an asymptotic momentum of the free electrons. This asymptotic photocurrent ultimately leads to THz radiation \cite{kim_terahertz_2007,sun_terahertz_2022,yu_anti-correlated_2022}. (2) The bandwidth of the THz radiation is directly proportional to the bandwidth of the pump pulse and inversely proportional to the duration of the pump pulse \cite{koulouklidis_spectral_2016}. 

In our previous study, we reported a novel bi-focal geometry involving two cascading plasmas, which revealed enhanced THz radiation with a bandwidth of up to 100 THz and a conversion efficiency of $\sim$0.1$\%$ \cite{zhang_intensity-surged_2022}. This phenomenon breaks the common-sense cognition that generating high-intensity super-continuum THz radiation requires pump pulses of longer wavelengths and wider bandwidths, as suggested by the photocurrent model. In this article, our aim is to exert experimental control over THz generation through the temporal-spatial manipulation of bi-focal bi-chromatic fields, thereby attempting to explain this counter-intuitive experimental phenomenon. The results demonstrate that temporal-spatial optimization of bi-chromatic fields can further increase THz radiation intensity. Our investigation involves analyzing the spectra of bi-chromatic fields and THz emissions, measuring the spatial location of the THz emission along the plasma filaments, and examining the dependence of THz intensity on laser intensity. These experimental results contradict the common assumptions of the photocurrent model. The present experimental findings, explored across multiple dimensions, pose significant challenges for theoretical investigations.

\section{Experimental methods}\label{sec2}

\begin{figure}[ht]%
\centering
\includegraphics[width=0.9\textwidth]{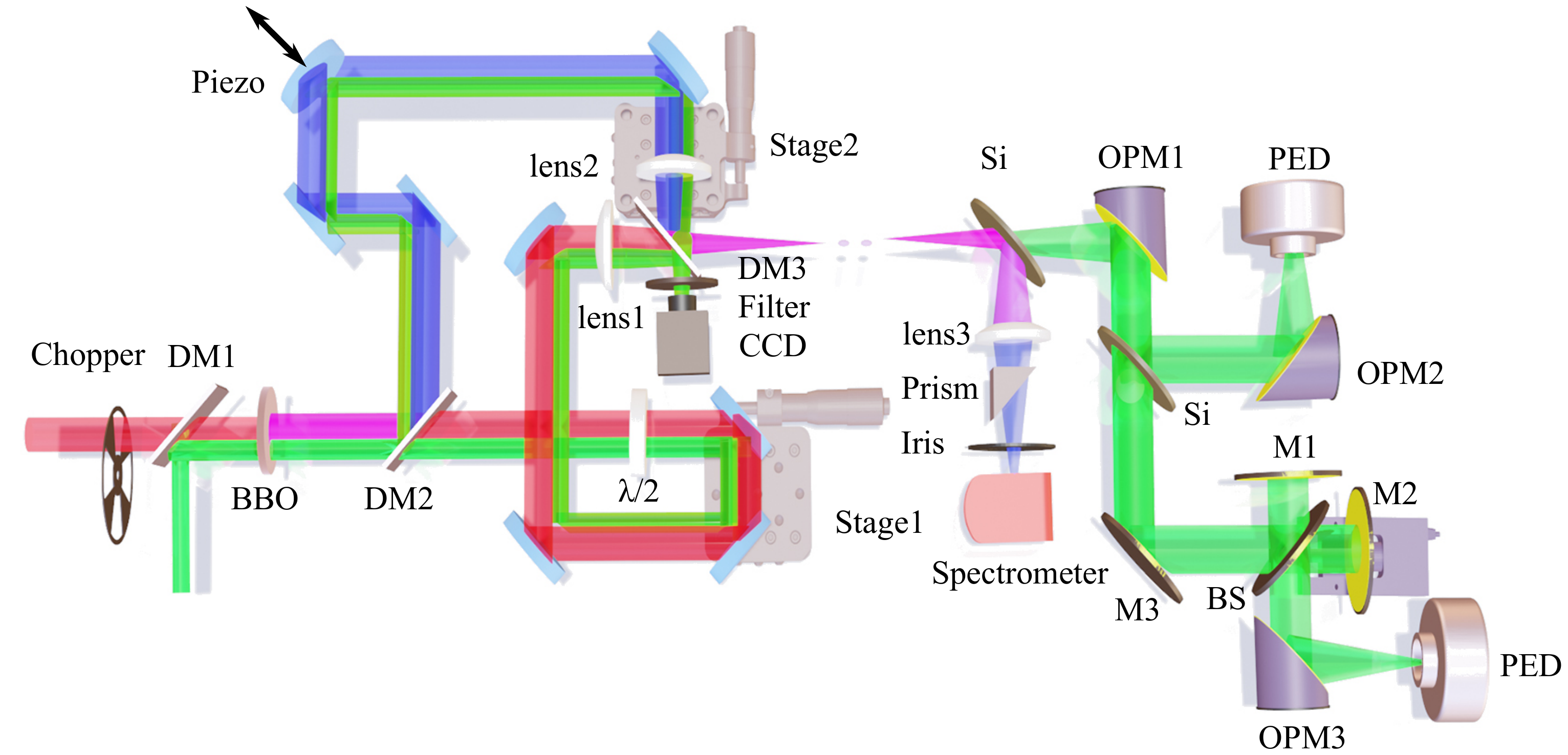}
\caption{Schematic illustration of the experimental setup: DM1-3, dichroic mirror; BBO, type-\uppercase\expandafter{\romannumeral1} $\beta$-barium borate crystal; Piezo, piezo stage; Si, silicon wafer; OPM1-3, off-axis parabolic mirror; BS, pellicle beam splitter; M1-3, THz mirror; PED, pyroelectric detector. The 3rd harmonic radiation is recorded by a spectrometer. The intensity and spectra of THz can be detected using the pyroelectric detector and a home-built Fourier transform spectrometer system.}\label{set up}
\end{figure}

In the experiment, we conducted joint measurements between THz generation and 3rd harmonic generation when linearly-polarized bi-chromatic fields are parallelly aligned. The experimental setup is depicted in Fig. \ref{set up}. A Ti:sapphire femtosecond amplification laser system delivers 40 fs (full width at half maximum (FWHM)) light pulses centered at $\omega$=810 nm with a maximal pulse energy of 1.7 mJ. The 2$\omega$ pulse is produced by the $\omega$ pulse after passing through a $\beta$-barium borate (BBO) crystal with a thickness of 200 $\mu$m and type-\uppercase\expandafter{\romannumeral1} phase-matching condition. This process yields an up-conversion efficiency of $\sim$30$\%$. Subsequently, the $\omega$-2$\omega$ fields are split by a dichroic mirror (DM2) into two arms of a Mach-Zehnder interferometer. In this setup, the time delay, relative orientation of polarization and focus conditions of $\omega$-2$\omega$ fields can be controlled, respectively. To suppress phase jitter between the $\omega$-2$\omega$ beams, an actively stabilized Mach-Zehnder interferometer is employed. This stabilization is achieved by introducing a continuous green laser (532 nm) that co-propagates with the $\omega$-2$\omega$ beams and generates interference. The interference fringes are monitored by a CCD camera, serving as a feedback signal. Real-time stabilization is facilitated by a mirror mounted on a piezo actuator that provides continuous feedback to maintain stable interference fringes. Upon stabilization, the relative phase fluctuation in the system remains below $0.02\pi$. After passing through the actively stabilized Mach-Zehnder interferometer, the $\omega$-2$\omega$ beams are combined by another dichroic mirror (DM3) to induce gas plasma ionization. 

The spatial manipulation of $\omega$-2$\omega$ pulses can be achieved by placing two lenses with a focal length of 100 mm (lens1, lens2) in each arm of the Mach-Zehnder interferometer. The lens for the 2$\omega$ pulse is mounted on a translation stage (Stage2), facilitating the fine-tuning of the spatial separation $d$ between the foci of $\omega$-2$\omega$ beams. The temporal separation $\tau$  of bi-chromatic pulses  can be practically controlled through two methods, including fine-tuning with sub-femtosecond accuracy and coarse-tuning with femtosecond accuracy. In the fine-tuning method, $\tau$ can be adjusted by introducing different co-propagating optical pathways for $\omega$-2$\omega$ beams due to the difference in refractive indices between the two pulses in the air. In addition, $\tau$ can also be coarsely changed by moving the piezo actuator in one arm of the interferometer.

The non-collinear propagation, resulting from the beam splitting and recombination of the bi-chromatic beams, disrupts the conical spatial distribution of the THz beam profile emitted along the forward direction. To address this misalignment issue, the THz beam profile is detected using a THz camera. In the Supplementary Note 1, the THz beam profiles are presented to showcase the successful collimation of the bi-chromatic beams at $d = 0\ \mathrm{mm}$, $d = 1\ \mathrm{mm}$ and $d = 2\ \mathrm{mm}$, respectively.

In the detection setup, the $\omega$ pulse and 2$\omega$ pulse after focusing as well as the generated third harmonic, are reflected by a polished silicon wafer and dispersed by a prism. The $\omega$ and 2$\omega$ beams are blocked by an iris serving as a spatial filter, allowing only the 3rd harmonic to couple into the fiber optic spectrometer (Thorlabs CCS200). Similarly, the $\omega$ and 2$\omega$ spectra are measured in the same manner. Subsequently, the transmitted THz is collimated and focused by two off-axis parabolic mirrors (OPM1, OPM2, 100 mm focal length) and detected by a pyroelectric detector (PED, THZ9B-BL Gentec-EO). During the measurement, the intensities of THz and 3rd harmonic can be jointly recorded as a function of $d$ and $\tau$ of $\omega$-2$\omega$ pulses.

The THz spectra are measured using a Fourier transform spectrometer based on the principle of a Michelson interferometer equipped with a 10 $\mu$m thick pellicle beam splitter (BS) and two THz mirrors (M1, M2). The THz waves passing through the Michelson interferometer are focused by a off-axis parabolic mirror (OPM3, 100 mm focal length) and detected by a pyroelectric detector. The Fourier transform spectrometer exhibits a flat response function above 15 THz. Wavelength calibration is conducted using an optical parametric amplifier with precisely defined wavelengths. Further details of the calibration process are presented in the Supplementary Note 2. 

\section{Results and Discussions}\label{sec3}

\subsection{THz waves and Third-order Harmonic Amplification via Temporal-spatial Manipulation}\label{subsec1}

\begin{figure}[ht]%
\centering
\includegraphics[width=1.0\textwidth]{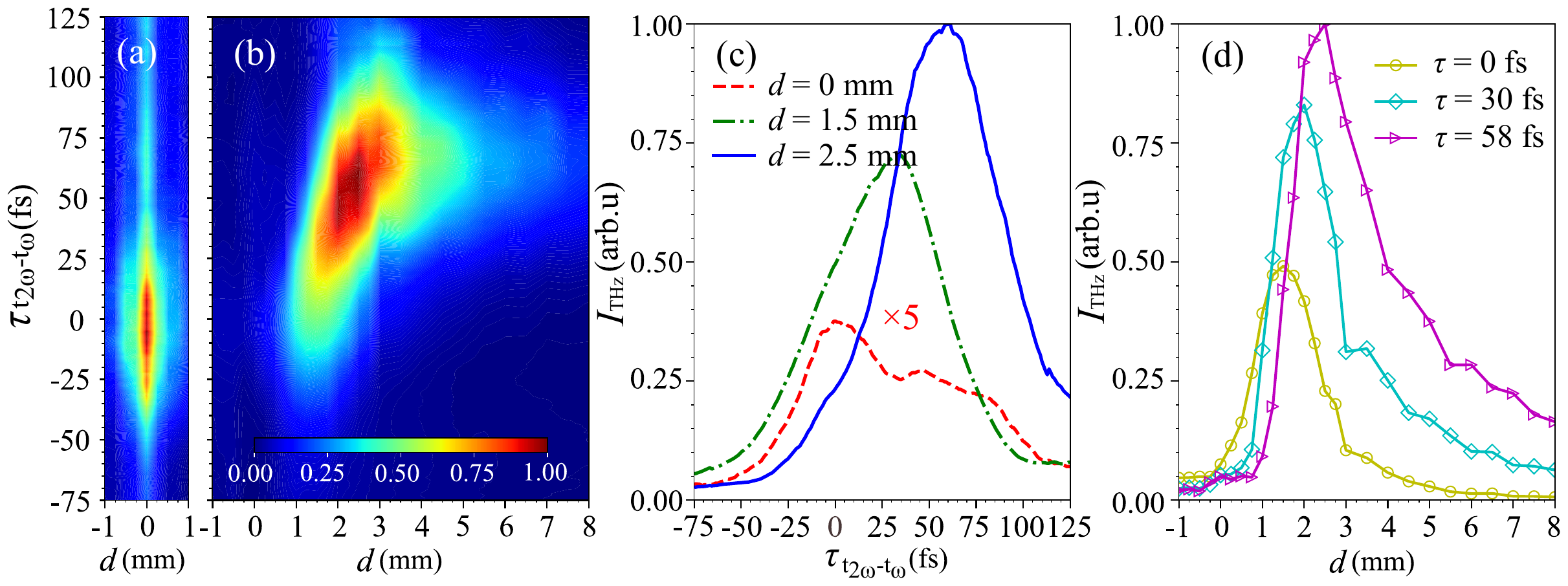}
\caption{Temporal and spatial modulation of THz waves and 3rd harmonic generation under bi-chromatic fields. (a) Measured 3rd harmonic intensity $I_{\mathrm{3rd}}(d,\tau)$ versus the $d$ and $\tau$. (b) The distribution of THz intensity $I_{\mathrm{THz}}(d,\tau)$ as a function of $d$ and $\tau$. (c) The THz intensities $I_{\mathrm{THz}}(\tau)$ projected from (b) at $d$ = 0 mm (red dashed line), $d$ = 1.5 mm (green dash doted line), and $d$ = 2.5 mm (blue solid line), respectively. (d) The THz intensities $I_{\mathrm{THz}}(d)$ projected from (b) at $\tau$ = 0 fs (yellow circles), $\tau$ = 30 fs (cyan diamonds) and $\tau$ = 58 fs (magenta right triangles), respectively.}\label{fig1}
\end{figure}

Fig. \ref{fig1}(a) depicts the 3rd harmonic intensity $I_{\mathrm{3rd}}(d,\tau)$ versus temporal-spatial manipulation and the calibration of the zero delay of $\omega$-2$\omega$ pulses. When referring to $\tau > 0\ \mathrm{fs}$, it indicates that the 2$\omega$ pulse temporally precedes the $\omega$ pulse. For $d > 0\ \mathrm{mm}$, it means that the plasma induced by the 2$\omega$ pulse is spatially located ahead of the plasma induced by the $\omega$ pulse along the laser propagation direction. The $\tau$ corresponding to the maximum 3rd harmonic yield is defined as the zero delay. $I_{\mathrm{THz}}(\tau)$ and $I_{\mathrm{3rd}}(\tau)$ with sub-femtosecond time delay, as well as the determination of the zero delay for the bi-chromatic fields specifically are provided in the Supplementary Note 3. $I_{\mathrm{THz}}(\tau)$ and $I_{\mathrm{3rd}}(\tau)$ exhibit anti-correlated behavior as a function of sub-femtosecond time delay, consistent with the previous measurements \cite{meng_phase_2023,fan_trajectory_2023}. 

In Fig. \ref{fig1}(b), the distribution of THz intensity $I_{\mathrm{THz}}(d,\tau)$ presents a lobe landscape  in the dimensions of spatial $d$ and temporal $\tau$ separations. Experimental THz intensity $I_{\mathrm{THz}}(d,\tau)$ sensitively depends on changes of $d$ and $\tau$, where a remarkable maximum distribution illuminates at certain $d$ and $\tau$ values. It reveals two counter-intuitive features that warrant attentions: (i) In the temporal dimension as shown in Fig. \ref{fig1}(c), the maximum THz generation appears when the bi-chromatic pulses are temporally separated. By scanning the time delay with femtosecond accuracy, we note that $I_{\mathrm{THz}}$ is insensitive versus $\tau$ at $d$ = 0 mm, where a broad distribution at about $\tau \approx$-25 fs and 100 fs is observed. However, peaked maxima of $I_{\mathrm{THz}}(\tau)$ at increasing time delays $\tau$ are more and more pronounced as $d$ increases and optimally $I_{\mathrm{THz}}$ with a remarkable 13 times high is achieved in Fig. \ref{fig1}(c) at $d$=2.5 mm and $\tau$=58 fs in comparison to that at $d$=0 mm and $\tau$=0 fs. (ii) In the spatial dimension as shown in Fig. \ref{fig1}(d), efficient optimization of THz generation occurs when the foci of bi-chromatic beams are noticeably separated. The maximum $I_{\mathrm{THz}}$ does not occur at $d = 0 \ \mathrm{mm}$, and $I_{\mathrm{THz}}$ is more efficient when $d > 0 \ \mathrm{mm}$ compared to $d < 0 \ \mathrm{mm}$. Analogy to the THz intensity dependence of two-pulse time delays shown in Fig. \ref{fig1}(c), peaked maximum distributions $I_{\mathrm{THz}}(d)$ shift to larger spatial separation $d$ as $\tau$ increases. 

These findings contrast with conventional expectations, where one would anticipate the THz yield to be optimized when the foci of the $\omega$-2$\omega$ beams are spatially and temporally overlapped. This observations are counter to the common expectation that the temporal and spatial overlapping of the bi-chromatic pulses would be a prerequisite for the most efficient THz generation and the $\omega$ laser
 is mainly responsible for the THz generation. According to the conventional understanding, both 3rd harmonic and THz radiation can be attributed to the Brunel radiation mechanism within the framework of the single-electron approximation \cite{brunel_harmonic_1990,zhang_continuum_2020,zhang_experimental_2020,he_third-order_2021}. For the case of present two-foci cascading plasma in time and in space, the most likely scenarios are as followings: The 2$\omega$ plasma filament plays a dominating role for THz generation (see experimental frequency spectrum later), where two color lasers are still prerequisites for highly efficient THz generation and the $\omega$ laser provides field modulation resulting in the symmetry breaking of a combined 2$\omega$-$\omega$ field.  To a large extent present two-foci cascading plasma configuration, where 2$\omega$ travels ahead in time and is focused ahead in space considering laser propagation direction, enable to avoid THz absorption by the $\omega$ created plasma. Theoretical calculations \cite{zhang_intensity-surged_2022} on the spatial modulation of bifocal field attributed to plasma absorption. In the present spatial geometry of two-foci, the $\omega$ laser field at focus spot of 2$\omega$ laser is relatively stronger only if $\tau>0$ in comparison to $\tau \leq 0$. That is to say that the optimal magnitudes of temporal and spatial separations for the highest THz emission will also be affected by laser profiles like intensity and pulse duration etc.. 

\subsection{ Spectral Measurements of Bi-chromatic Pulses}\label{subsec2}

\begin{figure}[ht]%
\centering
\includegraphics[width=1.0\textwidth]{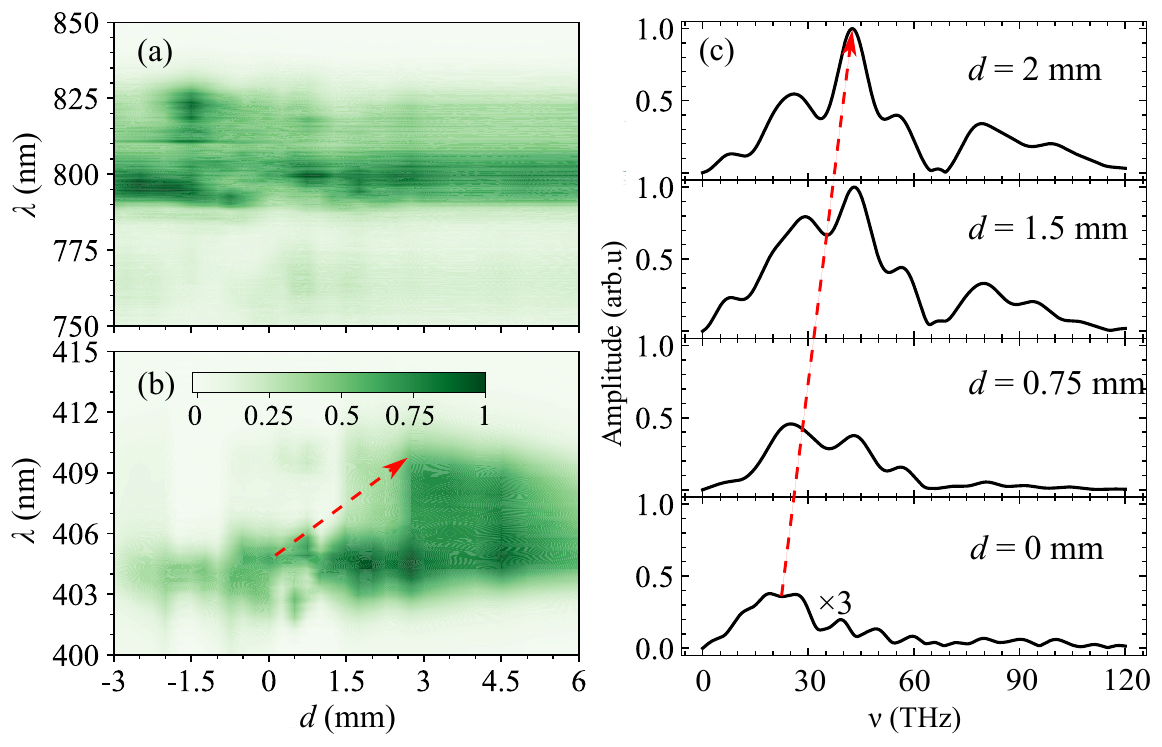}
\caption{The normalized spectra of $\omega$, 2$\omega$ and THz pulses. (a) The normalized spectra of $\omega$ emitted from plasma filament versus $d$. (b) The normalized spectra of 2$\omega$ emitted from plasma filament versus $d$, where the red arrow indicates the red shift of 2$\omega$ spectra. (c) The THz spectra for $d = 0\ \mathrm{mm}$, $d = 0.75\ \mathrm{mm}$, $d = 1.5\ \mathrm{mm}$ and $d = 2\ \mathrm{mm}$, respectively, where the red arrow indicates the blue shift of THz spectra. }\label{fig2}
\end{figure}

To gain insight into the origin of the enhancement of THz waves in the temporal-spatial manipulation of bi-chromatic fields, we measured the spectra of the bi-chromatic fields after focusing and the THz spectra emitted by the bi-chromatic fields at different $d$, which are shown in Fig. \ref{fig2}. The $\omega$ spectra versus $d$ are presented in Fig. \ref{fig2}(a), and we do not observe any shift in the $\omega$ spectra after focusing. Conversely, the red arrow in Fig. \ref{fig2}(b) indicates that the intensity of the 2$\omega$ spectra increases with $d$, along with a broadening on the red side of the 2$\omega$ spectra from $d$ = 0 mm to $d$ = 3 mm. It's worth noting that non-zero frequency detuning may occur naturally during the propagation of strong-field pulses. For example, the frequency shift due to plasma is blue shift, and the Kerr effect is red shift at the leading edge of the intensity \cite{gong_intensity-dependent_2021,huang_spatio-temporal_2022}. The $\omega$ and 2$\omega$ spectra are depicted as a function of $d$, illustrating that the $\omega$ pulse serves solely as a perturbation during THz generation process, It is observed that laser nonlinear propagation has a greater impact on the 2$\omega$ pulse compared to the $\omega$ pulse.

Furthermore, from the results of THz spectra versus $d$ in Fig. \ref{fig2}(c), it can be observed that the peak of THz spectra shifts from 20 THz to 45 THz as $d$ increases from 0 mm to 2 mm, as depicted by the red arrow. This blue shift is accompanied by a broadening of the THz spectra bandwidth by a factor of 6 compared to the $\omega$ spectra bandwidth. The red shift of the 2$\omega$ spectra and blue shift of the THz spectra demonstrate the energy transfer from the 2$\omega$ beam to the THz beam in the THz generation process. Whether based on the four-wave-mixing (FWM) process ($\omega$ + $\omega$ - 2$\omega$) or the photocurrent model, it is indicated that the broadening of 2$\omega$ spectra on the red side, driven by bi-chromatic laser fields, leads to a shift of the radiated THz pulses towards the high-frequency side \cite{babushkin_tailoring_2011,fan_tunable_2022}. Our experimental results are consistent with the theoretical description.

\subsection{Spatial Location of THz Radiation}\label{subsec3}

\begin{figure}[ht]%
\centering
\includegraphics[width=0.9\textwidth]{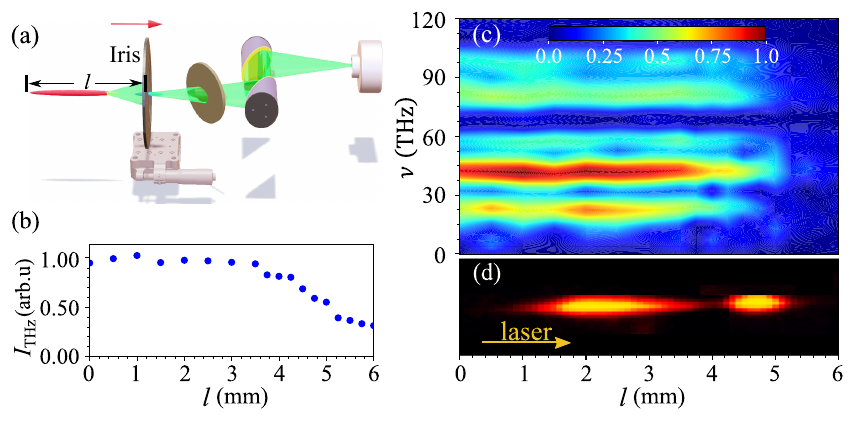}
\caption{The spatial location of THz emission on the plasma filament. (a) Measurement schematic of the spatial location of THz emission along the plasma filament. An iris with an aperture diameter of 0.5 mm is moved along the laser propagation axis, aligning with the plasma filament (depicted in red) passing through the aperture. As the iris is positioned, it effectively blocks the THz beam (depicted in green) emitted from the plasma filament on the left side of the iris. (b) THz intensity $I_{\mathrm{THz}}(l)$ as a function of the iris position $l$. (c) As the iris moves, THz spectra are measured for emissions originating from the plasma filament on the right side of the iris. (d) Snapshot of the fluorescence obtained by the CCD camera at $d = 2\ \mathrm{mm}$ and $\tau = 0\ \mathrm{fs}$.}\label{fig3}
\end{figure}

We conducted investigations on THz emission from various segments of the plasma filament, enabling us to pinpoint the spatial location of THz emission along the plasma filament, and gain valuable insights into the underlying mechanism driving THz amplification. The experimental schematic is shown in Fig. \ref{fig3}(a). An iris with an aperture diameter of 0.5 mm, concentric with the plasma filament, is moved along the propagation axis of the bi-chromatic beams. This manipulation results in blocking THz waves emitted from the plasma filament on the left side of the iris while enabling the detection of THz waves generated by the plasma filament on the right side of the iris. In this scenario, the plasma filament generated by the bi-chromatic fields is fixed in both spatial and temporal dimensions at $d$ = 2 mm and $\tau$ = 0 fs. The plasma fluorescence image is presented in Fig. \ref{fig3}(d).

The THz intensity $I_{\mathrm{THz}}(l)$ as a function of iris position $l$ is illustrated in Fig. \ref{fig3}(b), where $l$ = 0 mm is defined as the starting position of measurement. Notably, when the iris is positioned at the location of the plasma filament generated by the $\omega$ pulse, the detected THz intensity remains constant versus $l$. However, a sudden decrease in THz intensity is observed when blacking the THz emission from the $2\omega$ plasma filament, implying that the majority of THz pulses are generated from the $2\omega$ plasma filament rather than $\omega$ plasma filament. Here, we consider that the $\omega$ pulse plays an assisting role in the THz radiation process, and the propagation of the 2$\omega$ pulse is altered at the 2$\omega$ focus due to certain nonlinear effect, which in turn radiates THz pulses at the 2$\omega$ plasma position \cite{huang_shockwave-based_2023,hur_laser_2023}. Meanwhile, the THz spectra as a function of $l$ are measured using a Fourier transform spectrometer while translating the iris position $l$, and the results are illustrated in Fig. \ref{fig3}(c).  Employing the differential approach, the spectral distribution radiated from different segments of the plasma filament is presented in Supplementary Note 4. Notably, the high-frequency components of THz waves predominantly originate from the 2$\omega$ plasma filament, particularly the end of the 2$\omega$ plasma filament.

\subsection{Intensity-dependent Calibration}\label{subsec4}

\begin{figure}[ht]%
\centering
\includegraphics[width=0.5\textwidth]{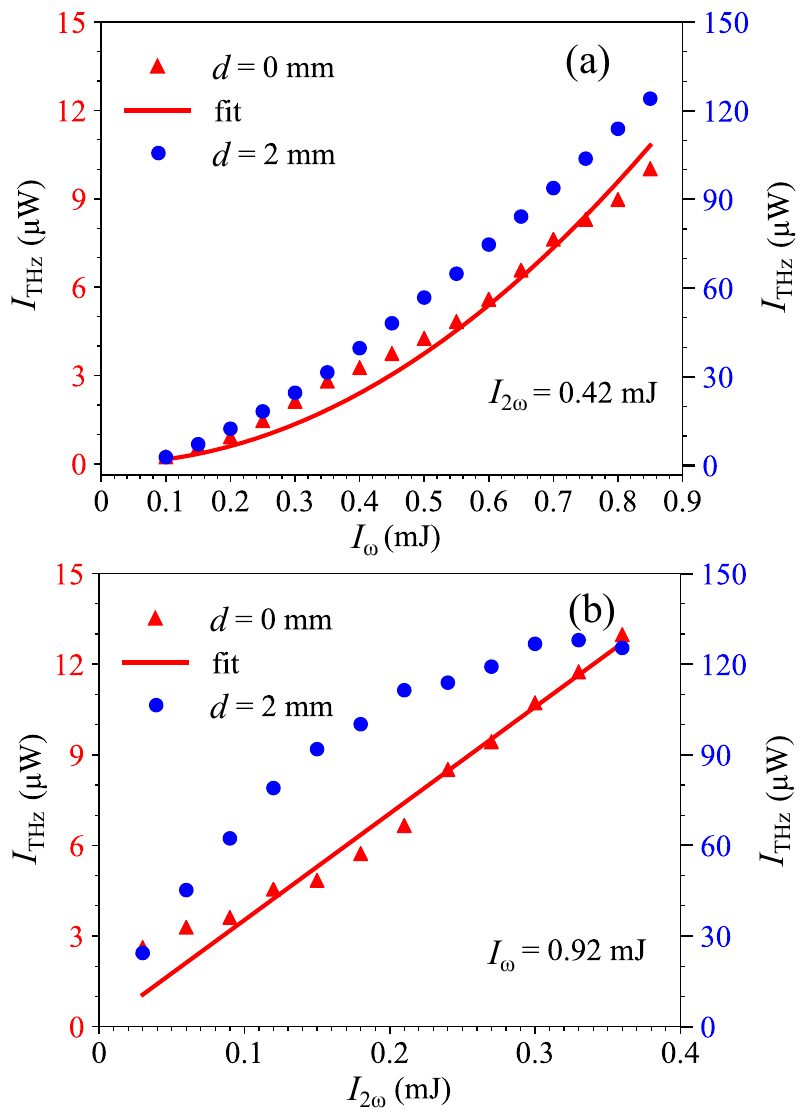}
\caption{Intensity-dependent calibration of THz yield. (a) The THz intensity $I_\mathrm{THz}$ is presented as a function of the $\omega$ pulse energy $I_{\omega}$, while keeping the 2$\omega$ pulse energy $I_{2\omega}$ constant at 0.42 mJ. (b) the $I_\mathrm{THz}$ is depicted as a function of $I_{2\omega}$, with the $I_{\omega}$ set at 0.92 mJ. The red triangles represent $I_\mathrm{THz}$ versus $I_{\omega}$ and $I_{2\omega}$ at $d$ = 0 mm and blue circles represent $I_\mathrm{THz}$ versus $I_{\omega}$ and $I_{2\omega}$ at $d$ = 2 mm, they correspond to the red and blue scales, respectively.}\label{fig4}
\end{figure}

Since our experimental arrangement permits us to adjust the energy of $\omega$ and 2$\omega$ beams individually, direct measurements of the emitted THz intensity $I_\mathrm{THz}$ versus $I_{\omega}$ at $d$ = 0 mm and $d$ = 2 mm, respectively, as depicted in Fig. \ref{fig4}(a). We also demonstrate the dependency of $I_\mathrm{THz}$ on $I_{2\omega}$ at $d$ = 0 mm and $d$ = 2 mm, respectively, as shown in Fig. \ref{fig4}(b). During the measurement, the energy of one beam is fixed while the energy of the other beam is changed. Typically, in the case of bifocal overlap ($d = 0\ \mathrm{mm}$) was reported previously \cite{kress_terahertz-pulse_2004,xie_coherent_2006,clough_laser_2012}: $\ensuremath{\boldsymbol{E}}_{\ensuremath{\operatorname{THz}}} \propto E_{\ensuremath{\operatorname{2\omega}}}{E_{\ensuremath{\operatorname{\omega}}}}^2$. 
Since the THz intensity is proportional to the square of the THz electric field, i.e. $\ensuremath{\boldsymbol{I}}_{\ensuremath{\operatorname{THz}}} \propto {\ensuremath{\boldsymbol{E}}_{\ensuremath{\operatorname{THz}}}}^2$, $\ensuremath{\boldsymbol{E}}_{\ensuremath{\operatorname{THz}}} \propto E_{\ensuremath{\operatorname{2\omega}}}{E_{\ensuremath{\operatorname{\omega}}}}^2$ can be rewritten as:
\begin{equation}
\ensuremath{\boldsymbol{I}}_{\ensuremath{\operatorname{THz}}} \propto I_{2\omega}{I_{\omega}}^2\label{Eq:intensity}
\end{equation} 
The red solid line in Fig. \ref{fig4}(a) and Fig. \ref{fig4}(b) represents the fitted results of $I_\mathrm{THz}$ versus $I_\mathrm{\omega}$ and $I_\mathrm{2\omega}$, respectively. These results are obtained from Eq. (\ref{Eq:intensity}) in combined with experimental data. This analysis corresponds to the case where $d$ = 0 mm. As predicted by the theory, the emitted $I_\mathrm{THz}$ is proportional to the square of $I_{\omega}$ (as observed in Fig. \ref{fig4}(a)) and to $I_{2\omega}$ above the ionization threshold (as observed in Fig. \ref{fig4}(b)). The observed dependence of $I_\mathrm{THz}$ on $I_{\omega}$ and $I_{2\omega}$ at $d = 0\ \mathrm{mm}$ is consistent with the conventional results.
 
Simultaneously, we measured the $I_\mathrm{THz}$ versus $I_{\omega}$ and $I_{2\omega}$ at $d = 2\ \mathrm{mm}$, respectively. The results demonstrate that $I_\mathrm{THz}$ is proportional to the square of the $I_{\omega}$ at $d$ = 2 mm, which is consistent with the trend observed at $d = 0\ \mathrm{mm}$. However, the relationship between $I_\mathrm{THz}$ and $I_{2\omega}$ shows a rapid increase followed by a slow increase, deviating from the trend at $d = 0\ \mathrm{mm}$. To confirm that this phenomenon is not exclusive to high laser intensities, we decreased the $\omega$ and 2$\omega$ laser intensity to explore the relationship between THz intensity and laser intensity. The observation reveals that the phenomenon persists under lower laser intensity (for more details, see the Supplementary Note 5). Thus, another solid piece of evidence is provided here that the increase in THz intensity at $d = 2\ \mathrm{mm}$ is due to the influence of the 2$\omega$ pulse.

Based on the aforementioned observations, it can be conclude that 2$\omega$ pulse plays a crucial role in both the enhancement and broadening of THz waves. 
An attempt was made to simulate the experiment using 3D propagation equations \cite{berge_3d_2013,tailliez_terahertz_2020}; however, a comprehensive elucidation of the experimental results cannot be achieved at present. As a result, a more thorough theoretical explanation is needed.

\section{Conclusion}\label{sec4}

In conclusion, our measurements reveal that THz generation is significantly optimized through the temporal-spatial manipulation of bi-chromatic fields. This optimization occurs when the two cascading foci are displaced, and temporal separation of the bi-chromatic pulses is introduced. By employing temporal and spatial modulation of the bi-focal bi-chromatic fields, the $I_{\mathrm{THz}}$ is amplified by a factor of 13 compared to the conventional bi-chromatic THz generation, while the bandwidth of THz extends 6 times beyond the pump light bandwidth. Meanwhile, detailed experiments have revealed counterintuitive phenomena: (1) Photon energy transfer from the 2$\omega$ beam to the THz beam is elucidated through 2$\omega$ and THz spectral correlation measurements; (2) The measurements of the location of THz emission from the plasma filament have revealed that the THz pulse originates from the 2$\omega$ plasma filament rather than the higher-electron-density $\omega$ plasma filament; (3) The investigation of the THz intensity dependence on the laser intensity reveals that when the two foci are pulled apart ($d$ = 2 mm), the dependence between $I_{\mathrm{THz}}$ and $I_{2\omega}$ clearly goes beyond the conventional scaling relationship. These intriguing and multi-perspective observations, which are beyond the prediction of the conventional photocurrent model, call for further theoretical investigations to offer comprehensive explanations and demonstrate potential applications in spectroscopic research.

\bmhead*{Supplementary information}

See supplementary information for additional information.

\bmhead*{Acknowledgments}

This work was supported by the National Key Research and Development Program of China (No. 2022YFA1604302) and the National Natural Science Foundation of China (No. 12334011, No. 12174284, and No.12374262). We extend our gratitude to Stefan Skupin from Claude Bernard University Lyon 1 for his valuable contributions and engaging discussions.

\bmhead*{Author contributions}

Experiments were designed by Jingjing Zhao, Yizhu Zhang and Yuhai Jiang. Jingjing Zhao conducted the experiments with assistance from Yizhu Zhang, Yanjun Gao, Meng Li, Xiaokun Liu. Weimin Liu provided mid-infrared sources to calibrate the THz Fourier transform spectrometer. Experimental data were analyzed and discussed by Jingjing Zhao, Yizhu Zhang, Tian-Min Yan and Yuhai Jiang. The manuscript was prepared by Jingjing Zhao, Yizhu Zhang and Yuhai Jiang after discussions and input from all authors.

\bmhead*{Competing interests}

The authors declare no competing interests.








\bibliography{main}

\end{document}


\subsection*{Supplementary Note 1: Alignment of bi-chromatic beams}
\renewcommand\thefigure{S\arabic{figure}}
\begin{figure}[htp]
  \centering
  \includegraphics[width=0.9\textwidth]{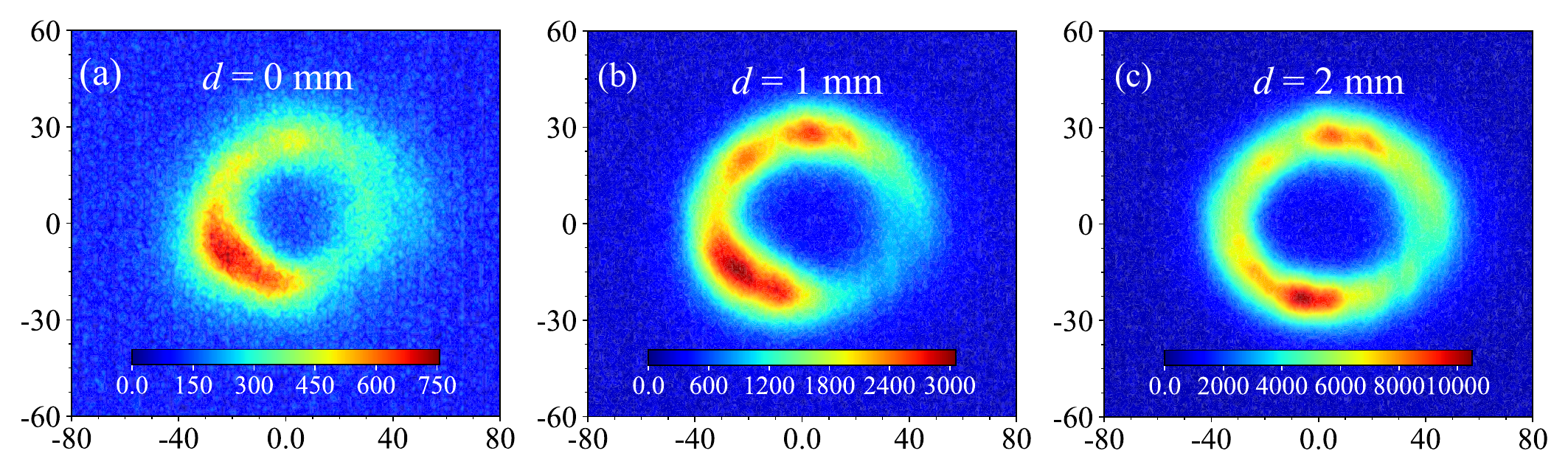}
  \caption{Spatial distributions of THz profiles at $d$ = 0 mm (a), $d$ = 1 mm (b) and $d$ = 2 mm (c), respectively.
  \label{fig:SI-1}}
\end{figure}

The beam profile of terahertz (THz) radiation can serve as a method to assess the alignment of bi-chromatic beams. When the bi-chromatic beams are well collimated and linearly propagate, the THz beam profile demonstrates a conical spatial distribution. However, any misalignment would disrupt this conical emission. Our measurements show that the THz waves cone emission remains consistent irrespective of the distance ($d$) of two foci (see Fig. \ref{fig:SI-1}), effectively ruling out the possibility that the enhancement of THz is a result of inappropriate collimation of the bi-chromatic beams.

\subsection*{Supplementary Note 2: Fourier transform spectrometer calibration}
\renewcommand\thefigure{S\arabic{figure}}
\begin{figure}[htp]
  \centering
  \includegraphics[width=0.6\textwidth]{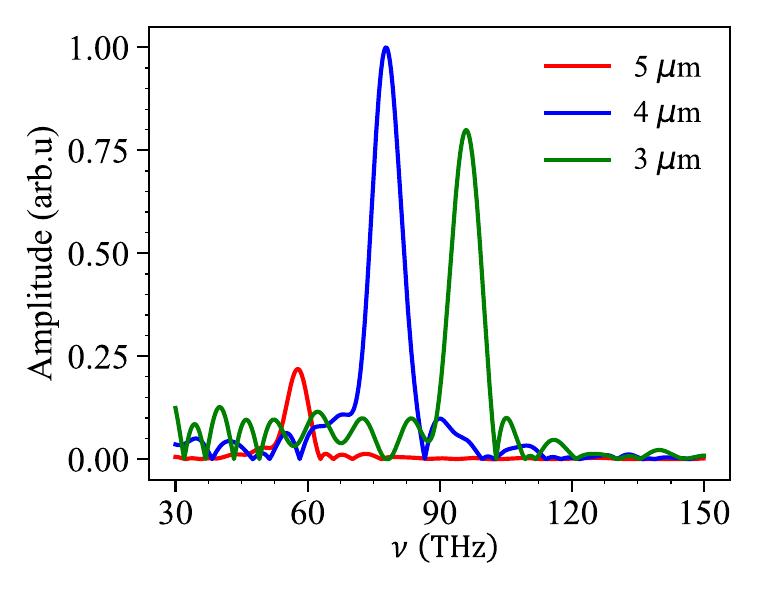}
  \caption{Calibration of the Fourier transform spectrometer involved obtaining spectra using incident laser fields with central wavelengths of 3 $\mu$m (green solid line), 4 $\mu$m (blue solid line) and 5 $\mu$m (red solid line), respectively.
  \label{fig:SI-2}}
\end{figure}

We calibrate the frequency axis of the home-built Fourier transform spectrometer to verify the accuracy of the bandwidth measurement. The wavelength of the femtosecond laser is tuned using an optical parametric amplifier (OPA, Coherent-OperA Solo). By setting the wavelength of OPA, three wavelengths of 3 $\mu$m (100 THz), 4 $\mu$m (75 THz) and 5 $\mu$m (60 THz), respectively, which are selected to calibrate the Fourier transform spectrometer. The spectra exhibits the same wavelength as set by the OPA, as shown in Fig. \ref{fig:SI-2}, confirming the accurate frequency measurement of the home-built Fourier transform spectrometer.

\subsection*{Supplementary Note 3: Joint measurement between 3rd harmonic and THz generation}
\renewcommand\thefigure{S\arabic{figure}}
\begin{figure}[htp]
  \centering
  \includegraphics[width=0.9\textwidth]{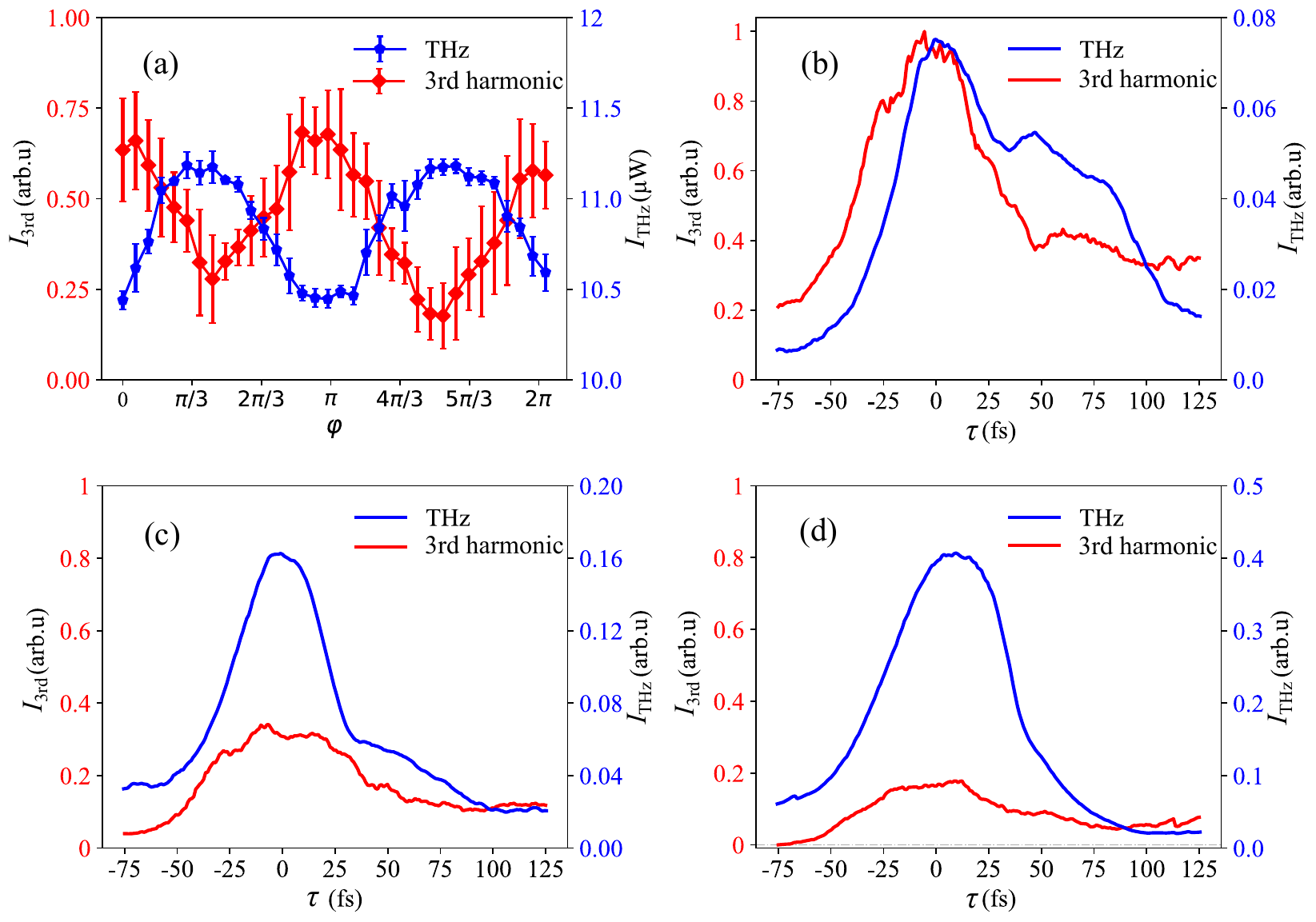}
  \caption{Joint measurement is conducted between 3rd harmonic and THz generation. (a) Demonstrates the anti-correlation between $I_{\mathrm{THz}}$ and $I_{\mathrm{3rd}}$ as a function of the relative phase ($\varphi$) of bi-chromatic pulses, with sub-femtosecond time delay when $d = 0$ mm. (b-d) The joint measurement between $I_{\mathrm{THz}}(\tau)$ and $I_{\mathrm{3rd}}(\tau)$ with femtosecond accuracy at $d = 0\ \mathrm{mm}$ (b), $d = 0.5\ \mathrm{mm}$ (c) and $d = 1.0\ \mathrm{mm}$ (d), respectively.
  \label{fig:SI-3}}
\end{figure}

The joint measurement between third-order harmonic (3rd harmonic) and THz generations is implemented to determine the absolute zero delay of bi-chromatic pulses. When the two plasma filaments are spatially overlapped ($d = 0\ \mathrm{mm}$), the intensity of 3rd harmonic and THz waves are measured as a function of the time delay, denoted as $I_{\mathrm{THz}}(\tau)$ and $I_{\mathrm{3rd}}(\tau)$, with sub-femtosecond time delay (relative phase ($\varphi$)) and with femtosecond accuracy, as shown in Fig. \ref{fig:SI-3} (a) and (b). Fig. \ref{fig:SI-3} (a) indicates that $I_{\mathrm{THz}}(\varphi)$ and $I_{\mathrm{3rd}}(\varphi)$ are anti-correlated as a function of the phase delay, while both $I_{\mathrm{THz}}(\tau)$ and $I_{\mathrm{3rd}}(\tau)$ exhibit coincidental maxima at $\tau = 0\ \mathrm{fs}$ (see Fig. \ref{fig:SI-3} (b)). These results are consistent with those obtained using the collinear geometry and are in accordance with conventional prediction.

As illustrated in Fig. \ref{fig:SI-3} (c) and (d), we observed that the optimal $\tau$ of 3rd harmonic generation remains unchanged when varying the separation of two plasma filaments $d$. However, the optimal $\tau$ of THz radiation shifts to $\tau > 0\ \mathrm{fs}$ when increasing $d$. By jointly measuring 3rd harmonic and THz radiation, we experimentally calibrated the zero delay $\tau = 0\ \mathrm{fs}$, enabling us to determine the optimal $\tau$ of THz radiation. This joint measurement of 3rd harmonic and THz waves confirms the conclusion that THz generation is optimized when the 2$\omega$ pulse temporally precedes the $\omega$ pulse.

\subsection*{Supplementary Note 4: Differential processing of THz spectra in plasma}
\renewcommand\thefigure{S\arabic{figure}}
\begin{figure}[htp]
  \centering
  \includegraphics[width=0.6\textwidth]{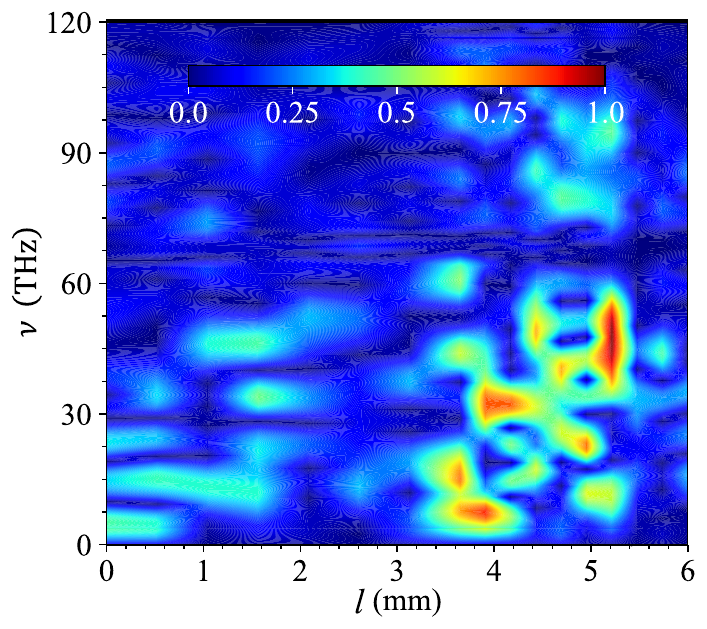}
  \caption{THz spectra are measured for emissions originating from various segments of the plasma filament.
  \label{fig:SI-4}}
\end{figure}

Building upon the results presented in Fig. 4(c) within the main text, we derived the THz spectra of the radiation from individual segments of the plasma filament through the application of differential processing. The results are shown in Fig. \ref{fig:SI-4}, it reveals that the high-frequency component of the THz spectra primarily originates from the tail of the second harmonic (2$\omega$) plasma filament

\subsection*{Supplementary Note 5: Dependence of THz yield on laser intensities}

In the Fig. 5 of the main text, when $\tau$ = 0 fs, the THz yield exhibits distinct intensity-dependent behaviors at $d = 0\ \mathrm{mm}$ and $d = 2\ \mathrm{mm}$, respectively. We further investigate this relationship by selecting different $\omega$ and $2\omega$ intensities to study the dependence of THz yield on $I_{2\omega}$ and $I_{\omega}$, as shown in Fig. \ref{fig:SI-5}. The results in Fig. \ref{fig:SI-5} are consistent with those in the Fig. 5 of the main text, indicating that the dependence of the THz yield on the laser intensities does not vary with the laser intensity.
\renewcommand\thefigure{S\arabic{figure}}
\begin{figure}[htp]
  \centering
  \includegraphics[width=0.9\textwidth]{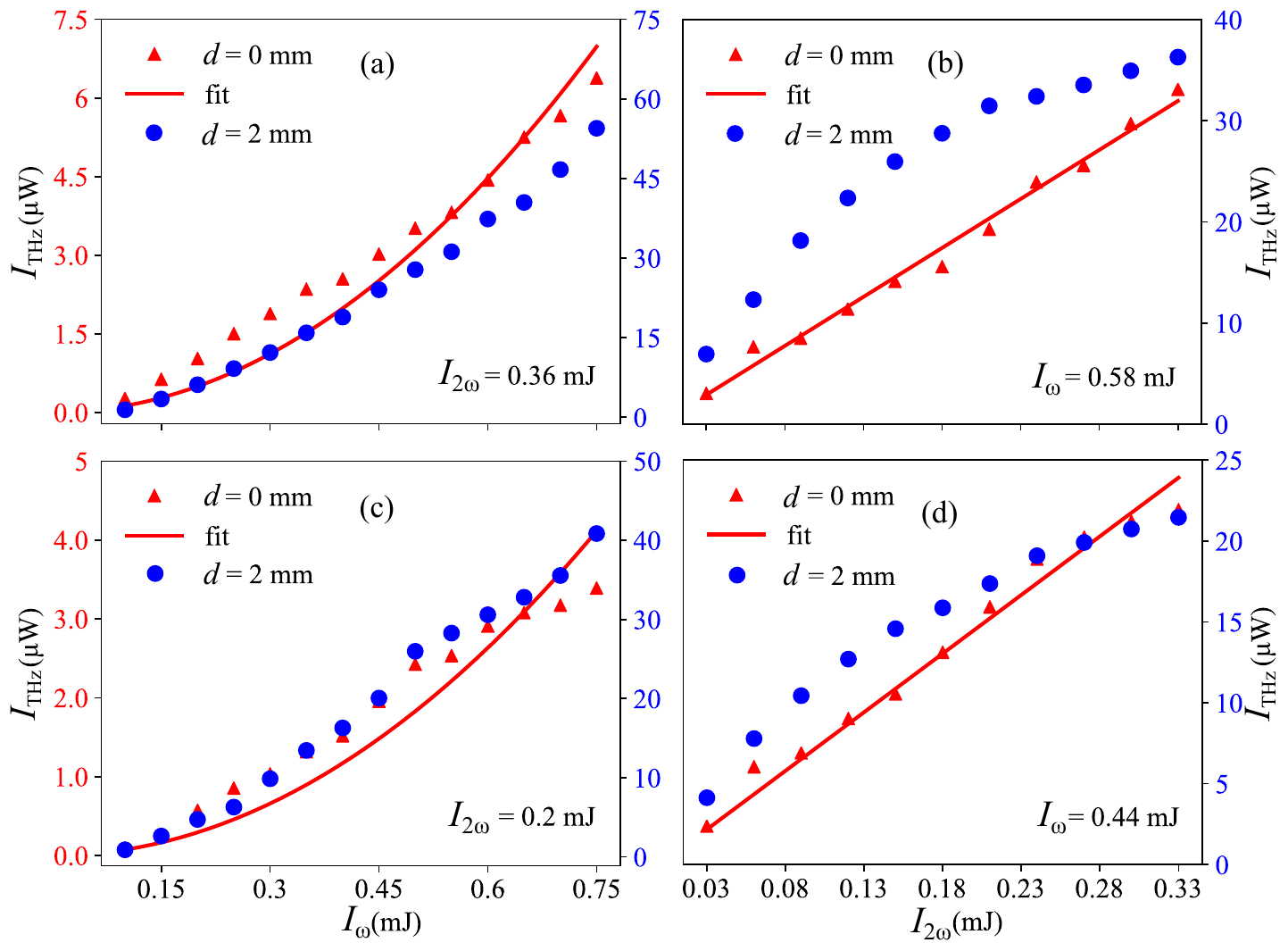}
  \caption{(a) and (c) Dependence of THz yield on the $\omega$-field intensities; (b) and (d) Dependence of THz yield on 2$\omega$-field intensities.
  \label{fig:SI-5}}
\end{figure}